# Akhmediev breather signatures from dispersive propagation of a periodically phase-modulated continuous wave

Ugo Andral [1], Bertrand Kibler [1], John M. Dudley [2], and Christophe Finot [1,*]

[1] *Laboratoire Interdisciplinaire Carnot de Bourgogne, UMR 6303 CNRS-Université de Bourgogne-Franche-Comté, 9 avenue Alain Savary, BP 47870, 21078 Dijon Cedex, France*

[2] *Institut FEMTO-ST, CNRS Université de Bourgogne Franche-Comté UMR 6174, 25030 Besançon, France*

[*] *Corresponding author:*

*E-mail address: christophe.finot@u-bourgogne.fr*

*Tel.: +33 3 80395926*

**Abstract:** We investigate in detail the qualitative similarities between the pulse localization characteristics observed using sinusoidal phase modulation during linear propagation and those seen during the evolution of Akhmediev breathers during propagation in a system governed by the nonlinear Schrödinger equation. The profiles obtained at the point of maximum focusing indeed present very close temporal and spectral features. If the respective linear and nonlinear longitudinal evolutions of those profiles are similar in the vicinity of the point of maximum focusing, they may diverge significantly for longer propagation distance. Our analysis and numerical simulations are confirmed by experiments performed in optical fiber.

# I. Introduction

The study of solitons on finite background, and more specifically the class of Akhmediev breather (AB) [1] structures is a subject that is attracting much interest in the context of studying nonlinear localization in extreme events and modulation instability processes. In particular, in a nonlinear medium governed by the nonlinear Schrödinger equation (NLSE) with focusing nonlinearity, a small periodic perturbation (typically a sinusoidal amplitude or intensity modulation) imposed on a continuous background, can display rapid growth and decay characteristics which have been interpreted as prototype signatures of extreme or rogue wave dynamics [2]. Examples of such breather solutions have been demonstrated in hydrodynamics [3,4], plasma physics [5,6] and in optics [7-14]. Indeed, with progress in ultrafast measurement techniques, experiments performed in optical fiber have proven to be an especially convenient means of exciting nonlinear AB solutions and studying their localization properties in great detail [7-10]. In addition, the properties of AB solutions at their point of maximum localization (temporal compression) have motivated studies of applications in high repetition optical source development [11,12] and provided insights into the physics of supercontinuum formation [13].

A seemingly unrelated area of research that has also received much recent attention is the use of sinusoidal phase modulation on a continuous wave field for high speed temporal signal processing and other applications in ultrafast optics. For example, when combined with a dispersive element and taking advantage of the powerful time/space analogy [15,16], such phase modulation has enabled the development of a range of ultrafast sources [17,18], as well as the implementation of novel concepts such as lenticular lenses for optical sampling [19]. Very recently, the quality of pulse generation using this technique has been improved using simple additional triangular spectral phase shaping that enables elimination of unwanted sidelobes and background [20].

Interestingly, the pulse localization characteristics observed using sinusoidal phase modulation and those seen during the evolution of AB solutions possess a number of qualitative similarities. However, to our knowledge, these have not been the subject of any previous studies. Our objective here is to investigate these similarities in more detail, where we initially consider the characteristics of the localization profiles obtained at the point of maximum focusing, before then studying the longitudinal evolution of those profiles. Our analysis and numerical simulations are then confirmed by experiments performed in optical fiber.

# II. Breathers and phase-modulated pulses at their point of maximum compression

We first consider an Akhmediev-breather structure (AB) propagating in an optical fiber with Kerr nonlinear coefficient $\gamma$ and second order anomalous dispersion $\beta_2$ ($\beta_2 < 0$). The spatio-temporal evolution of the complex field envelope of the AB in the slowly varying envelope approximation (and neglecting a term of propagating phase) is analytically described by [2] :

$$\psi_B(t,\xi_B) = \sqrt{P_0}\,\frac{(1-4a)\cosh(b\,\xi_B) + i\,b\sinh(b\,\xi_B) + \sqrt{2a}\cos(\omega_m\,t)}{\sqrt{2a}\cos(\omega_m\,t) - \cosh(b\,\xi_B)} \qquad (1)$$

The normalized distance $\xi_B$ is related to the dimensional distance via $\xi_B = z/L_{NL}$ with the nonlinear length $L_{NL} = 1/(\gamma P_0)$ for an initial plane wave of power $P_0$. The parameters $a$ and $b$ are the normalized perturbation frequency and instability growth rate, and in dimensional units are given by $2a = 1-(\omega_m/\omega_c)^2$ and $b = \sqrt{8\,a\,(1-2a)}$. Here $t$ is the temporal coordinate and $\omega_m$ denotes the dimensional modulation frequency in the range $0 < \omega_m < \omega_c$, where $\omega_c^2 = 4\gamma P_0/|\beta_2|$. Here $0 < a < 0.5$ with the limiting value $a = 0.5$ leading to the well-known Peregrine soliton solution [21]. The temporal period is $T_0 = 2\pi/\omega_m$ and at maximum compression, i.e. $\xi_B = 0$, Eq. (1) reduces to:

$$\psi_B(t,\xi_B = 0) = \sqrt{P_0}\,\frac{(1-4a) + \sqrt{2a}\cos(\omega_m\,t)}{\sqrt{2a}\cos(\omega_m\,t) - 1} \qquad (2)$$

In the frequency domain, the AB spectrum [i.e., the Fourier transform $s(\omega)$ of the temporal field $\psi(t)$] consists of a discrete comb of spectral lines $s_{Bn} = s(n\,\omega_m)$ that are equally spaced by $\omega_m$ and with an amplitude given by [8,12,22] (here, factors of constant amplitude and phase are ignored) :

$$\begin{cases} s_{B0}(\xi_B) = \dfrac{i\,b\sinh(b\,\xi_B) + p^2\cosh(b\,\xi_B)}{\sqrt{\cosh^2(b\,\xi_B) - 2a}} - 1 \\ s_{Bn}(\xi_B) = \dfrac{i\,b\sinh(b\,\xi_B) + p^2\cosh(b\,\xi_B)}{\sqrt{\cosh^2(b\,\xi_B) - 2a}} \left[ \dfrac{\cosh(b\,\xi_B) - \sqrt{\cosh^2(b\,\xi_B) - 2a}}{\sqrt{2a}} \right]^{|n|} \end{cases} \quad (3)$$

with $n \geq 1$ and $p = \sqrt{2}\,\omega_m / \omega_c$. By writing:

$$\begin{cases} r = \dfrac{i\,b\sinh(b\,\xi_B) + p^2\cosh(b\,\xi_B)}{\sqrt{\cosh^2(b\,\xi_B) - 2a}} \\ q = \dfrac{\cosh(b\,\xi_B) - \sqrt{\cosh^2(b\,\xi_B) - 2a}}{\sqrt{2a}} \end{cases}, \quad (4)$$

one can rewrite Eq. (4) in the following form:

$$\begin{cases} s_{B0} = r - 1 \\ s_{Bn} = r\,q^{|n|} \end{cases} \quad (5)$$

At the point of maximum compression, the expressions for $r$ and $q$ reduce to $r = 2\sqrt{1-2a}$ and $q = \left(1 - \sqrt{1-2a}\right)\big/\sqrt{2a}$ ), so that $s_{Bn}$ are real. We plot in Fig. 1 the temporal and spectral intensity profiles obtained for $a = 0.2$ (panels a1 and b1 respectively, black lines).

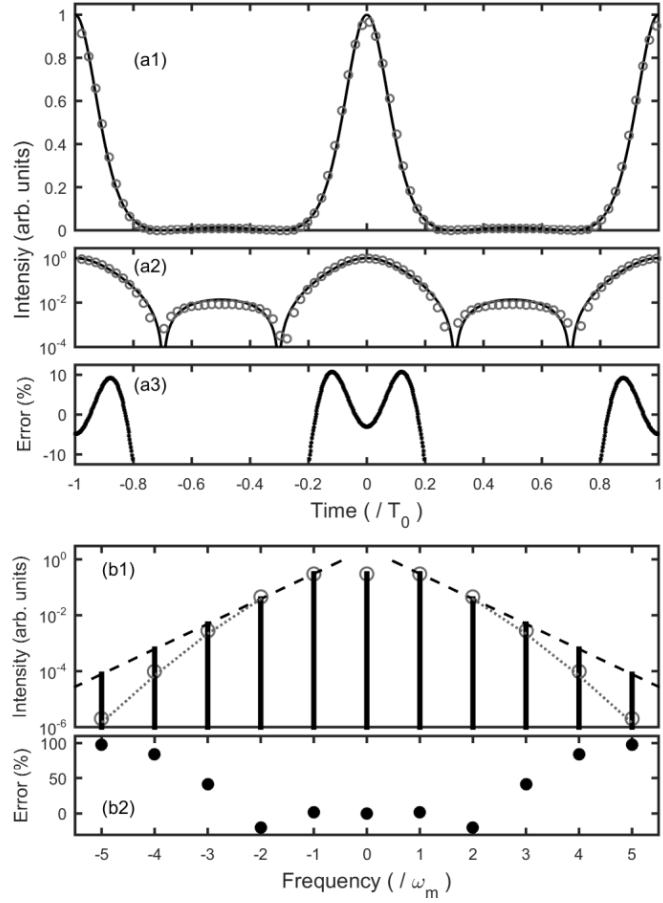

**Figure 1 :** Comparison of **(a)** temporal and **(b)** spectral properties of Akhmediev breather with $a = 0.2$ (solid black lines, Eq. (2) and (5)) and a phase modulated sinusoidal wave with $A = 1.43$ rad (open circles, Eq. (8)) at the point of maximum compression for the same modulation frequency $\omega_m$. Temporal profiles are compared on linear and logarithmic scales (panels a1 and a2, respectively) and panel (a3) represents the local error $E_T$ between the two waveforms when the intensity is significant (above 5% of the peak power). The black dashed and the grey dotted lines in panel (b1) represent the evolution of the spectral wings of the waveforms as predicted by Eq. (10) and (11). Local error $E_S$ when comparing the two spectra is plotted on panel (b2).

Turning now to the case of a phase-modulated field, we consider the properties of a continuous wave whose temporal phase has been sinusoidally modulated at the same angular frequency $\omega_m$ as the AB solution above:

$$\psi_{P0}(t) = e^{i\, A \cos(\omega_m t)} \tag{6}$$

with $A$ the amplitude of the phase modulation. Using a Jacobi-Angi expansion, Eq. (6) can be rewritten as a sum of sinusoidal functions [23]:

$$\psi_{P0}(t) = \sum_{n=-\infty}^{\infty} i^n J_n(A)\, e^{i n \omega_m t} \tag{7}$$

leading to a comb spectrum made of components $s_{Pn}$ at frequency $n\,\omega_m$ (with respect to the central frequency) having an amplitude of $s_{Pn} = J_n(A)$, where $J_n$ is the Bessel function of the first kind of order $n$. The wave is therefore not Fourier limited. For $A < 2.4$, applying a triangular spectral phase profile enables the generation of a Fourier transform signal [20] that can be expressed as:

$$\psi_P(t) = J_0(A) + 2\sum_{n=1}^{\infty} J_n(A) \cos(n\,\omega_m t) \tag{8}$$

In order to compare the properties of the AB with a continuous phase modulated wave, we have computed the misfit factors $M_T$ and $M_S$ regarding the temporal and spectral profiles defined as:

$$\begin{cases} M_T^2 = \int (I_{T,B} - I_{T,P})^2\, dt\, /\, \int I_{T,B}^2\, dt \\ M_S^2 = \sum_n (I_{S,Bn} - I_{S,Pn})^2\, /\, \sum_n I_{S,Bn}^2 \end{cases} \tag{9}$$

with $I_{T,x}(t)$ and $I_{Sn,x}$ the temporal and spectral intensity profiles obtained for the breather (subscript $x = B$) and phase modulated signal (subscript $x = P$) and normalized to have the same average power. For $a = 0.2$, $M_T$ and $M_S$ can reach values as low as 0.044 and 0.025 when $A = 1.43$ rad. Temporal and spectral profiles of the wave obtained at the point of maximum compression for this value are plotted with circles in Fig. 1. We can clearly see that the two profiles are nearly undiscernible. We also plot on Fig. 1(a3) the temporal local error defined as $E_T(t) = (I_{T,P} - I_{T,B}) / I_{T,B}$ and we can see that this local remains well below 10% when the pulse intensity is significant, i.e. above 5% of the peak power. Regarding the spectrum, Fig. 1(b2) reveals that if the $E_S(n\,\omega_m) = (I_{S,Pn} - I_{S,Bn}) / I_{S,Bn}$ is extremely low in the central part of the spectrum, some discrepancy appears in the wings. Indeed, the AB has wings characterized by an exponential decrease of its tails leading to a typical triangular signature when plotted on a logarithmic scale [8,13,22] (see black dashed line in Fig. 1(b1)). The ratio between two successive components of the AB (with $n \geq 1$) is indeed fixed:

$$\frac{I_{F,Bn+1}}{I_{F,Bn}} = \left|\frac{s_{Bn+1}}{s_{Bn}}\right|^2 = q^2 \tag{10}$$

On the contrary, for moderate values of $A$, it can be shown that the spectral wings of $\psi_P$ follow a different trend (see grey dotted lines in Fig. 1(b1)) :

$$\frac{I_{F,Pn+1}}{I_{F,Pn}} = \left|\frac{s_{Pn+1}}{s_{Pn}}\right|^2 = \frac{A}{4(n+1)^2} \tag{11}$$

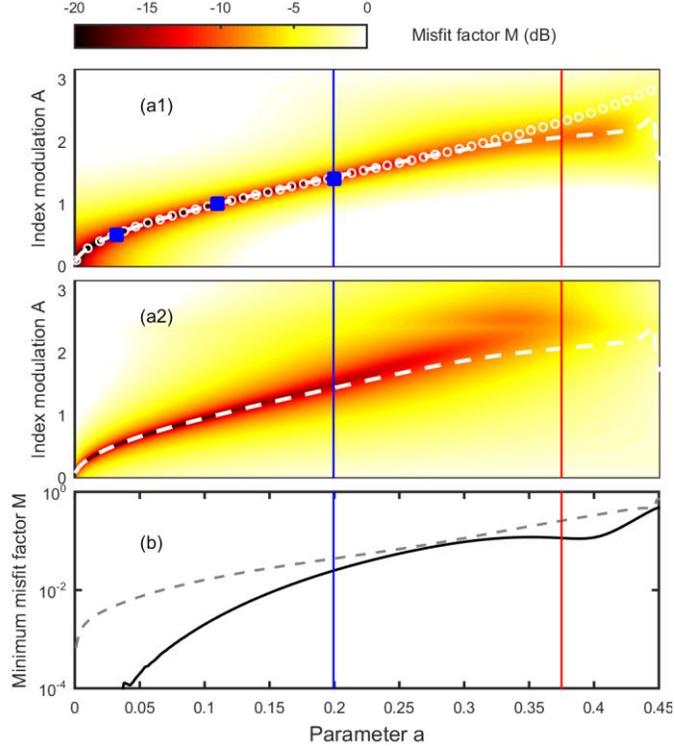

**Figure 2 :** **(a)** Evolution of the spectral (a1) and temporal (a2) misfit factor $M_S$ and $M_T$ as a function of the parameter $a$ of the AB (horizontal axis) and the index of modulation $A$ of the sinusoidal phase (vertical axis). The red and blue vertical lines represent the value $a = 3/8$ and the value of $a$ used for the figures in this theoretical section, respectively. The white dashed lines represent $A_0$, i.e. the optimum value of $A$ that achieves, for a given $a$, the minimal spectral misfit factor. The open circles represent the theoretical predictions derived from Eq. (12). The blue squares represent the points used for the experimental validation presented in section 4) **(b)** Summary of the spectral and temporal misfits factors (black solid and dashed grey lines, respectively) obtained for $A_0$.

A more exhaustive study showing the evolution of the misfit factor between the AB and a wave resulting from an initial phase modulation is shown in Fig. 2 for both the temporal and spectral intensity profiles. We note that higher $a$ parameters lead to higher misfit factors between the two waves at their point of maximum compression. The value $a = a_c = 3/8$ appears as a turning point, leading to a pronounced increase in the misfit. Indeed, for $a > a_c$ and as can be derived from Eq. (5), the central component of the spectrum, i.e. the continuous background of the wave, becomes in opposite of phase with respect to the rest of the spectrum. For $a < a_c$, the misfit factors that can be reached remains low, below 0.2, indicating temporal and spectral profiles that are extremely close. As can be seen from panels (a) of Fig. 2, $A$ has to be carefully chosen because the tolerance for obtaining low values of $M$ is quite narrow. The optimum value $A_0$ to be applied to fit the spectrum obtained for a given value of $a$ is plotted with a white dashed line. We can therefore make out that achieving low $M_S$ values also leads to low $M_T$ values. Increasing values of $a$ corresponds to increasing values of $A_0$. One can also notice that, for low values of $a$, $A_0$ is in remarkable agreement with the following guideline:

$$A_0 = 4\,q \tag{12}$$

# III. Evolution of breather and phase-modulated continuous wave upon propagation

Let us now consider the longitudinal evolution of the two waves. The longitudinal evolutions of the temporal and spectral properties of breathers in the presence of nonlinearity and dispersion are provided by Eq. (1) and (5), respectively. Regarding the linear evolution of the wave given by Eq. (8), the propagation in a dispersive medium induces a quadratic phase on the spectrum, leading to an additional spectral phase on the component $s_n$ of $\exp\left(i\,\beta_2\,z\,(n\,\omega_m)^2/2\right)$. This leads in the temporal domain to:

$$\psi_P(t,\xi_P) = J_0(A) + 2\sum_{n=1}^{\infty} J_n(A)\left(\cos\left(n^2\,\xi_P/2\right) + i\sin\left(n^2\,\xi_P/2\right)\right)\cos\left(n\,\omega_m\,t\right) \tag{13}$$

where $\xi_P = \beta_2\,\omega_m^2\,z$ is a normalized distance that takes into account both the level of dispersion of the fiber and the frequency of modulation.

Figure 3 compares the evolution of the temporal and spectral intensity profiles of the AB and the phase-modulated wave that propagates in a nonlinear and linear medium, respectively, in the case of $a = 0.2$ and $A = 1.43$ rad. We can therefore notice that in the vicinity of the maximum of compression, the longitudinal evolutions of the two waves seem rather close. Indeed, both waves experience a stage of temporal compression followed by a symmetrical stage where the peak power of the structures decreases. In both cases, the wave seems to tend towards a continuous intensity profiles. Note however that the longitudinal scales that are involved are rather different. Inspection of the evolution of the spectrum emphasizes one of the intrinsic major differences between the two waves : whereas the nonlinear evolution of the AB manifests itself with major changes in the spectrum with a growth and return cycle [8], the linear propagation does not affect the spectral intensity profile of the wave.

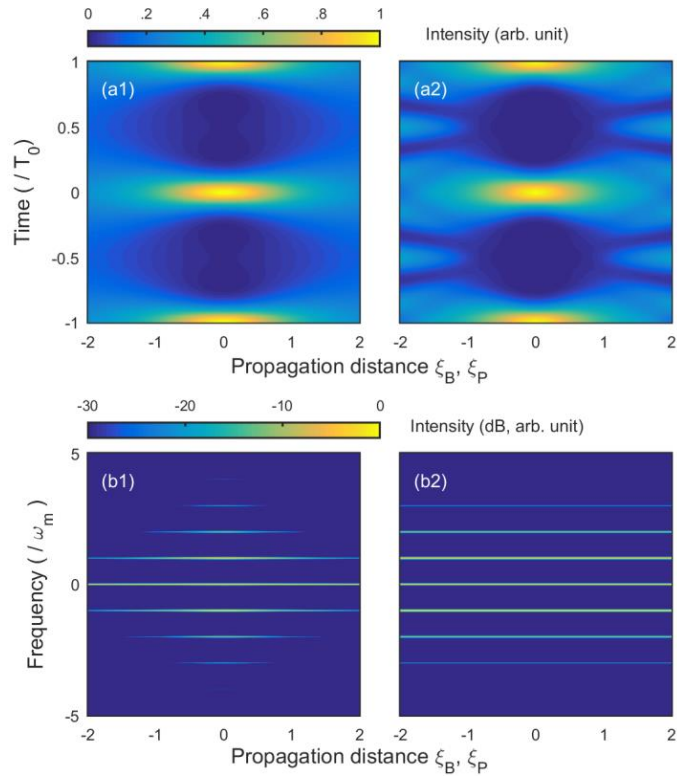

**Figure 3 :** Comparison of the longitudinal nonlinear propagation of an AB with $a = 0.2$ (panels 1) with the linear evolution of a tailored sinusoidal phase modulated wave with $A = 1.43$ rad (panels 2). The temporal and spectral evolutions are shown in panels (a) and (b) respectively.

Details of the evolution of the peak-power of the structures are reported on Fig. 4 and exhibit very similar trends in the vicinity of the point of maximum compression. However, let us recall that the length scales are different.

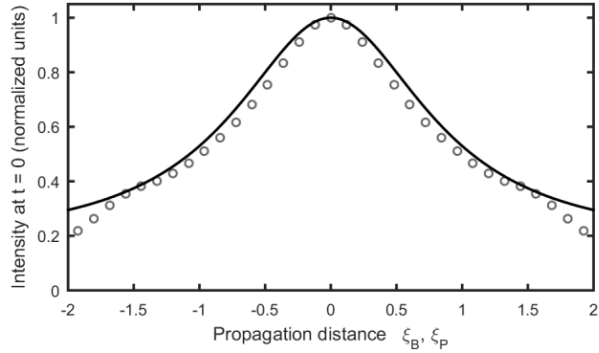

**Figure 4 :** Comparison of the evolution of the peak power of the structure for the nonlinear propagation of an AB (solid black line) and for the linear propagation of a phase modulated continuous wave (open grey circles).

When observed over a longer propagation distance (Fig. 5), differences between the temporal evolution of the two waves become much more pronounced. Indeed, a perfect AB grows from a continuous wave and then decays without leaving any trace, a manifestation of the celebrated Fermi-Pasta-Ulam recurrence [24]. On the contrary, the phase-modulated wave never disappears and a carpet pattern appears, with same structure repeating with temporal phase shift of half a period. Similar patterns resulting from the diffraction of periodic objects are well-known in the field of spatial optics as the Talbot effect [25] and this self-imaging process can be extended to the temporal domain [26,27] using the very rich time/space analogy [15,28]. It is worth noting that those typical phase shifts can be observed in the nonlinear propagation of a breathers when some losses are included [9,29].

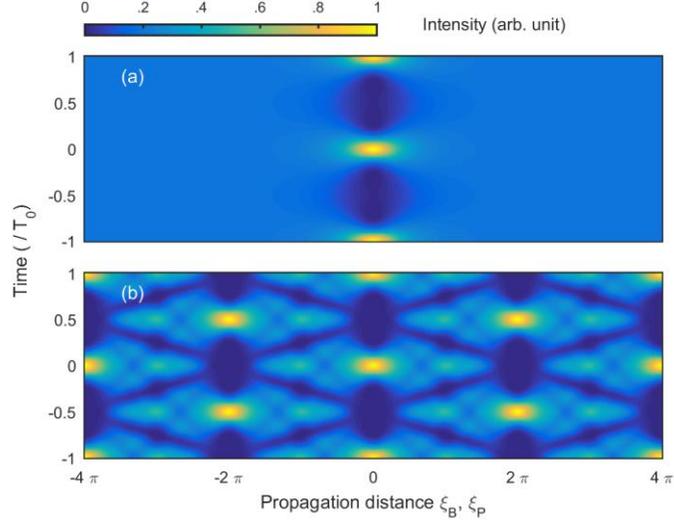

**Figure 5 :** Longitudinal evolution of the temporal intensity profiles of an AB with $a = 0.2$ (a) and of a sinusoidally phase modulated signal with $A = 1.43$ rad (b). Results of Fig. 3(a) are here plotted on a longer propagation distance in order to emphasize the Talbot carpet.

A further means to visualize the various similarities and differences between the two waves is to plot their evolution in terms of the power and phase of their main spectral components. Indeed, we represent the evolution on Fig. 6 of the parameters $\eta$ and $\phi$ [30] defined respectively for this problem with spectra having an even symmetry as $\eta = |s_0|^2 / \left(|s_0|^2 + 2|s_1|^2\right)$ and $\phi = 2\arg(s_1) - \arg(s_0)$. We can note that with these coordinates, the AB describes a separatrix. On the contrary, the trajectory of the phase modulated pulse is a full circle. Both trajectories are similar only in a small region around the maximum of compression, i.e. when the central component interferes constructively with the harmonic modulation.

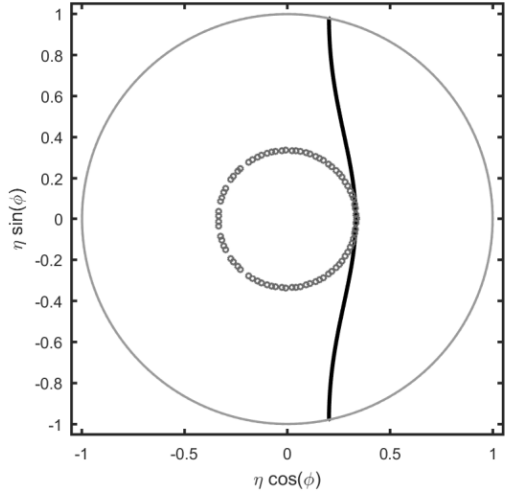

**Figure 6 :** Evolution of the spectral properties of the AB with *a* = 0.2 and phase modulated signal with *A* = 1.43 rad in a nonlinear and linear propagation respectively (solid black curve and open circles respectively).

# IV. Experimental results

In order to validate our analytical and numerical studies above, we have implemented the experimental setup is sketched on Fig. 7 and is based on devices that are commercially available and typical of the telecommunication industry. A continuous wave laser at 1550 nm is first temporally phase modulated using a Lithium Niobate electro-optic device driven by an amplified sinusoidal electrical signal. A linear spectral shaper (Finisar Waveshaper) based on liquid crystal on silicon technology is then used to apply the spectral triangular phase pattern required to generate Fourier-transform limited structures [31]. We operate at a repetition rate of 20 GHz. In order to ensure enhanced environmental stability, polarization-maintaining components have been used. The resulting signal is directly recorded by means of a high-speed optical sampling oscilloscope (1 ps resolution) and with a high-resolution optical spectrum analyser (5 MHz of resolution).

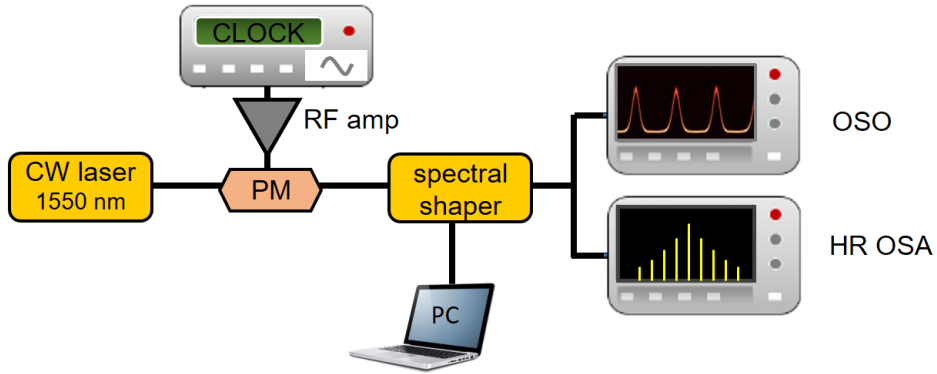

**Figure 7 :** Experimental setup.

Temporal and spectral intensity profiles synthesized at the point of maximum compression for different values of *A* are summarized of Fig. 8. We can note an excellent agreement between theses profiles resulting from the phase modulation of a continuous wave and the typical profiles of breathers. The typical values *a* of those breathers that fits the experimental data are plotted on Fig. 2 with squares and are in very good agreement with our analysis above.

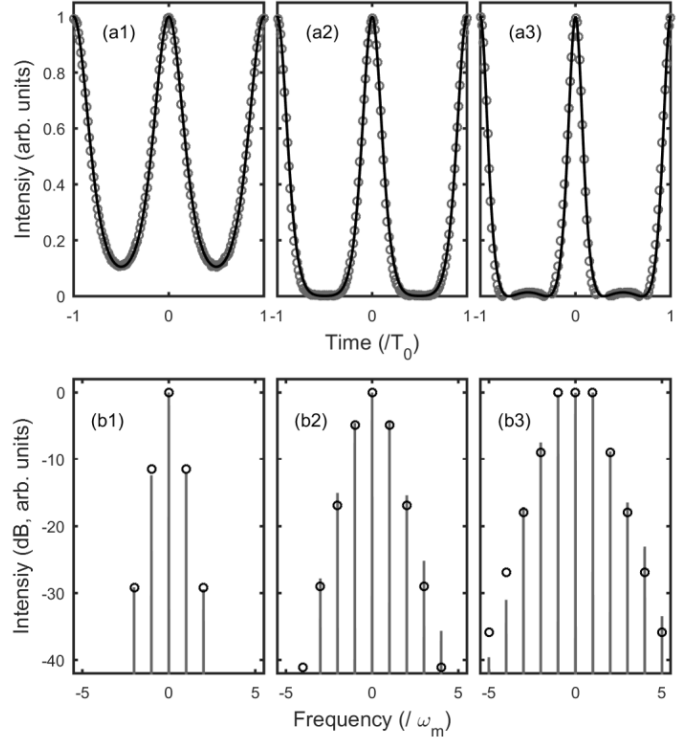

**Figure 8 :** Temporal and spectral intensity profiles (panels a and b, respectively). Experimental records obtained for phase modulation $A$ of 0.5, 1 and 1.41 rad (grey circles or line ; panels 1, 2 and 3, respectively) are compared with the profiles of a AB with $a$ parameters of 0.032, 0.11 and 0.2 (black line or circles).

By superimposing a quadratic spectral phase to the pulse achieved at the point of maximum compression, we are able to simulate the linear propagation of the structure [31]. Acting on the sign of the parabolic spectral phase formally enables us to mimic propagation in negative and positive values. As can be seen on Fig. 9(a), the experimental evolution of the intensity profile obtained for $A = 1.41$ rad can be qualitatively reproduced in the vicinity of the point of maximum compression by the longitudinal evolution of an AB with $a = 0.2$. When we move away for the compression point, discrepancies increase.

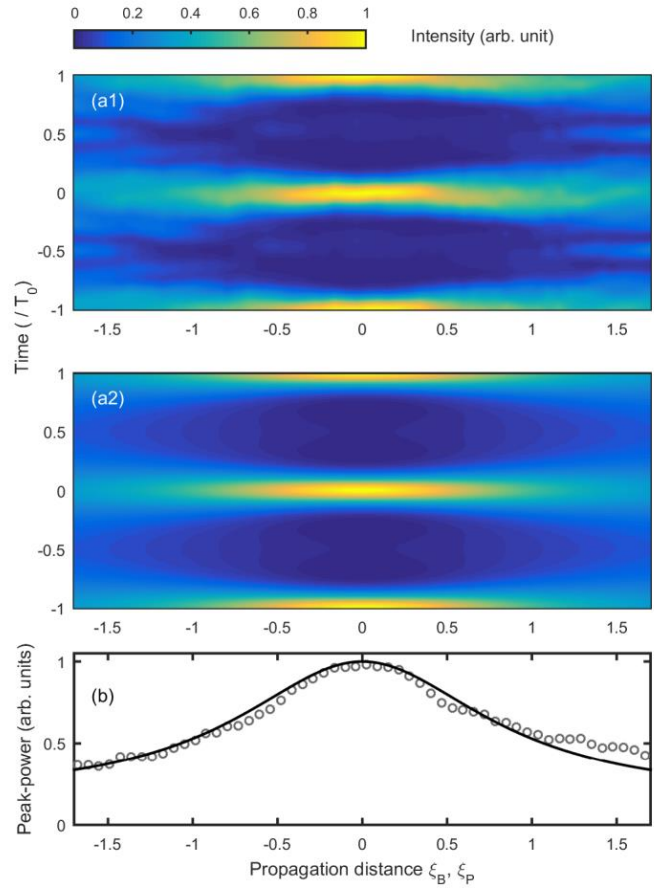

**Figure 9 :** Longitudinal evolution of the temporal intensity. (a) Experimental records for a phase-modulated signal with $A = 1.4$ rad. (panel a1) are compared with the theoretical evolution of an AB for $a = 0.2$ (panel a2). (b) Longitudinal evolution of the peak-power for the phase-modulated signal (grey circles) and for the AB (black line).

# V. Conclusions

In conclusion, we have theoretically and numerically shown that a continuous wave with a periodic phase modulation evolving in a linear medium and an Akhmediev breather governed by the NLSE may present similar signatures. Indeed, for typical parameter *a* below 3/8, their temporal and spectral profiles at the point of maximum compression can be very close. Moreover, they both experience similar longitudinal evolution of their temporal properties in the vicinity of the maximum compression distance. Such unexpected similarities have been confirmed experimentally.

These results highlight that care should be devoted when identifying localized structures in a chaotic field and attributing their origin a priori to nonlinear propagation [32,33]: the longitudinal evolution of a temporal structure shape is not sufficient to clearly address its linear or nonlinear nature. Knowledge of the powers that are involved is crucial and differences between the linear and nonlinear structures can also be apparent in the frequency domain. As breathers have also been stressed in other systems based on the Lugiato-Lefever equation , same care should also hold for the analysis of structures emerging from a cavity.

This study also indicates new ways to easily generate profiles that are close to Akhmediev breathers. Indeed, previous methods may have relied on the spectral shaping of a comb [34] or on the use of an initial sinusoidally modulated intensity profile [8,11] associated with further propagation in a nonlinear fiber, as discussed in details in [14] and [35]. Here, we show that a simple and cost-efficient alternative based on electrooptic phase modulation yields AB-like structures with low *a* in the vicinity of the point of maximum compression. It has moreover the advantage of avoiding detrimental consequences of potential Brillouin backscattering.

# Acknowledgements

We acknowledge the support of the Institut Universitaire de France (IUF), the Bourgogne-Franche Comté Region, the French Investissements d'Avenir program and the Agence Nationale de la Recherche (ISITE-BFC ANR-15-IDEX-0003 Projects Bright and Nextlight).

# References

[1]   N. N. Akhmediev and V. I. Korneev, Theor. Math. Phys. **69**, 1089 (1986).

[2]   N. Akhmediev, A. Ankiewicz, and M. Taki, Phys. Lett. A **373**, 675 (2009).

[3]   A. Chabchoub, B. Kibler, J. M. Dudley, and N. Akhmediev, Phil. Trans. R. Soc. A **372**, 20140005 (2014).

[4]   M. Onorato, D. Proment, G. Clauss, and M. Klein, Plos One **8**, e54629 (2013).

[5]   H. Bailung, S. K. Sharma, and Y. Nakamura, Phys. Rev. Lett. **107**, 255005 (2011).

[6]   S. A. El-Tantawy, A. M. Wazwaz, and S. Ali Shan, Physics of Plasmas **24**, 022105 (2017).

[7]   K. Hammani, B. Kibler, C. Finot, P. Morin, J. Fatome, J. M. Dudley, and G. Millot, Opt. Lett. **36**, 112 (2011).

[8]   K. Hammani, B. Wetzel, B. Kibler, J. Fatome, C. Finot, G. Millot, N. Akhmediev, and J. M. Dudley, Opt. Lett. **36**, 2140 (2011).

[9]   C. Naveau, P. Szriftgiser, A. Kudlinski, M. Conforti, S. Trillo, and A. Mussot, Optics Letters **44**, 763 (2019).

[10]  A. Tikan *et al.*, Phys. Rev. Lett **119**, 033901 (2017).

[11]  J. Fatome, B. Kibler, and C. Finot, Opt. Lett. **38**, 1663 (2013).

[12]  B. Varlot, Y. K. Chembo, and C. Finot, Microw. Opt. Technol. Lett. **56**, 664 (2014).

[13]  J. M. Dudley, G. Genty, F. Dias, B. Kibler, and N. Akhmediev, Opt. Express **17**, 21497 (2009).

[14]  B. Kibler, in *Shaping Light in Nonlinear Optical Fibers*, edited by S. Boscolo, and C. Finot (John Wiley & Sons Inc, 2017).

[15]  V. Torres-Company, J. Lancis, and P. Andrés, in *Progress in Optics*, edited by E. Wolf (Elsevier, 2011), pp. 1.

[16]  C. Froehly, A. Lacourt, and J. C. Viénot, Nouvelle Revue d'Optique **4**, 183 (1973).

[17]  T. Kobayashi, H. Yao, K. Amano, Y. Fukushima, A. Morimoto, and T. Sueta, IEEE J. Quantum Electron. **24**, 382 (1988).

[18]  T. Komukai, Y. Yamamoto, and S. Kawanishi, IEEE Photon. Technol. Lett. **17**, 1746 (2005).

[19]  J. Nuno, C. Finot, and J. Fatome, Opt. Fiber Technol. **36**, 125 (2017).

[20]  U. Andral, J. Fatome, B. Kibler, and C. Finot, Submitted, arXiv:1904.01875 (2019).

[21]   B. Kibler, J. Fatome, C. Finot, G. Millot, F. Dias, G. Genty, N. Akhmediev, and J. M. Dudley, Nature Physics **6**, 790 (2010).

[22]   N. Akhmediev, A. Ankiewicz, J. M. Soto-Crespo, and J. M. Dudley, Phys. Lett. A **375**, 775 (2011).

[23]   M. Abramowitz and I. A. Stegun, *Handbook of mathematical functions: with formulas, graphs, and mathematical tables* (Courier Corporation, 1964), Vol. 55.

[24]   N. Devine, A. Ankiewicz, G. Genty, J. Dudley, and N. Akhmediev, Phys. Lett. A  (2011).

[25]   J. Wen, Y. Zhang, and M. Xiao, Adv. Opt. Photon. **5**, 83 (2013).

[26]   T. Jannson and J. Jannson, J. Opt. Soc. Am. **71**, 1373 (1981).

[27]   J. Azaña and L. R. Chen, J. Opt. Soc. Am. B **20**, 1447 (2003).

[28]   R. Salem, M. A. Foster, and A. L. Gaeta, Adv. Opt. Photon. **5**, 274 (2013).

[29]   O. Kimmoun *et al.*, Sci. Rep. **6**, 28516 (2016).

[30]   S. Trillo and S. Wabnitz, Opt. Lett. **16**, 986 (1991).

[31]   M. A. F. Roelens, S. Frisken, J. Bolger, D. Abakounov, G. Baxter, S. Poole, and B. J. Eggleton, J. Lightw. Technol. **26**, 73 (2008).

[32]   M. Närhi *et al.*, Nat Commun **7**, 13675 (2016).

[33]   S. Toenger, T. Godin, C. Billet, F. Dias, M. Erkintalo, G. Genty, and J. M. Dudley, Scientific Reports **5**, 10380 (2015).

[34]   B. Frisquet, B. Kibler, and G. Millot, Phys. Rev. X **3**, 041032 (2013).

[35]   M. Erkintalo, G. Genty, B. Wetzel, and J. M. Dudley, Phys. Lett. A **375**, 2029 (2011).